# Dissecting the Spatial Structure of Cities from Human Mobility Patterns to Define Functional Urban Boundaries

September 2017

Francisco Humeres & Horacio Samaniego[*]

Laboratorio de Ecoinformática, Instituto de Conservación, Biodiversidad y Territorio, Universidad Austral de Chile, Campus Isla Teja s/n, Valdivia, Los Ríos, Chile

*Corresponding author Email: horacio@ecoinformatica.cl

**Abstract**

Since the industrial revolution, accelerated urban growth has overflown administrative divisions, merged cities into large built extensions, and blurred the boundaries between urban and rural land-uses. These traits, present in most of contemporary metropolis, complicate the definition of cities, a crucial issue considering that objective and comparable metrics are the basic inputs needed for the planning and design of sustainable urban environments. In this context, city definitions that respond to administrative or political criteria usually overlook human dynamics, a key factor that could help to make cities comparable across the urban fabric of diverse social, cultural and economic realities.

Using a technique based on the spectral analysis of complex networks, we rank places in 11 of the major Chilean urban regions from a high-resolution human mobility dataset: Official origin-destination (OD) surveys. We propose a method for further distinguishing urban and rural land-uses within these regions, by means of a network centrality measure from which we construct a spectre of geographic places. This spectre, constructed from the ranking of locations as measured by their approximate number of embedded human flows, allows us to probe several urban boundaries. From the analysis of the urban scaling exponent of trips in relation to the population across these city delineations, we identify two clearly distinct scaling regimes occurring in urban and rural areas. The comparison of our results with land cover derived from remote sensing suggests that, for the case of trips, the scaling exponent in urban areas is close to linear. We conclude with estimations for well-formed cities in the Chilean urban system, which according to our analysis could emerge from clusters composed by places that capture at least ~138 trips (over the expectation) of the underlying mobility network.

**INTRODUCTION**

At the beginning of the 20$^{th}$ century, the biologist Patrick Geddes coined the neologism "conurbations" for describing the physical merging of human settlements that were emerging as by-products of the industrial revolution [1]. Today most of the world's largest conurbations [2,3], are characterized by a rapid expansion into the surrounding hinterland in the form of low-density urbanization [4,5], "leap-frog" or scattered development, and new centralities that often compete economically with the urban core [6]. As a by-product of such growth pattern, urban areas have extended beyond their original administrative boundaries, while urban and rural land-uses have become blurred, making the definition of cities a highly problematic issue.

The problem of defining cities can be understood as part of a longstanding problem in spatial analysis referred to as the *modifiable area unit problem*, where incorrect choices for spatial units of aggregation can lead to inconsistent results and misleading conclusions [7–9]. This is a particularly sensitive issue for urban data, as aggregates vary widely according to the city definition employed. The extent of the problem is such that the United Nations World Urbanization Prospects report warns its readers of the inconsistency of city definitions across the countries included in the document [10].

The lack of reliable urban estimates is often associated to significant controversies in urban planning. Particularly in urban transportation and policy research, where the link between the impact of urban expansion and environmental sustainability has been a hotly debated topic [11–13]. More recently, critics have raised concerns about the alarming figures of urban growth given the lack of a standardized definition for cities [14]. Some scholars have even argued that such figures have provided ideological support for megaprojects with dubious claims in terms of economic efficiency and environmental sustainability [15–17].

The issue of city definition has also come to the fore among the recent explosion of quantitative approaches to study and describe urban systems[18]. A central aspect of such description, neatly summarized in Batty's book [19], seeks to provide a new framework to account for the extensively observed nonlinear scaling relationship between urban functions and city size[20,21]. In fact, Pumain in 2004 and Bettencourt and colleagues in 2007, inspired by recent explanations for allometric scaling in biology [22,23], expose the relation of most urban functions using the relationship $Y = Y_o N^\beta$ [21], which describes the association of most urban indicators *Y*, with the size of the city *N*, where $Y_o$ is a constant scale factor and *β* the scaling exponent, also termed elasticity in the economics literature.

Empirical evidence from cities across the world provide ample support for the existence of urban scaling [24–28]. Nevertheless, conflicting results have been reached regarding the nature of urban scaling of other urban outputs as $CO_2$ emissions [29–33] for instance. A problem that Barthelemy and Louf partly attribute to the variety of city definitions used [29,34]. Recent exploratory analysis of the British and French urban systems conducted by Arcaute, Batty et. al [8,35], show the sensitivity of scaling exponents to the city definition employed.

We here develop a method for delineating cities from human activity patterns, drawing from the analysis of the network of common places visited by people. The approach offers the potential to define urban areas with independence from administrative or political boundaries and is unconstrained by criteria of spatial concentration such as population density or the contiguity of built-up areas.

By iteratively exploring the parameter space leading to different city realizations, we develop a methodology capable to separate a general area into clusters akin to *urban* and *rural* areas. For each of these realizations we fit a power law, finding two different scaling regimes for urban areas and its segregated rural hinterland with respects to human mobility.

**Cities as the overlap of human activities**

We seek to define the functional limits of cities as derived from the mobility of their inhabitants rather than from indicators of spatial concentration. While we base this analysis on commuting areas, we envisage that this tool can easily be extended to employ novel

Information and Communication Technology (ICT) datasets with the added advantage of real-time updates, and reduced collection costs [36].

**Spectral Techniques and Socio-Spatial Place Ranking**

Commonly used spectral techniques applied in complex network analyses are used to provide efficient ranking of geographic nodes in large mobility networks. Recent applications of such approach include the analysis of massive and high-resolution geolocated data produced by ubiquitous sensors embedded in mobile and hand-held devices risen from the advent of ICTs. Specific examples are the use of dimensional reduction to infer temporal patterns of human activity [37] and the ranking of places according to human interactions [38]. Others have studied mobility networks including neighborhood detection from social media data [39], clustering of human trip behavior [40] using the various implementations of Newman's modularity algorithm [41] and the geographic segmentation of mobility data among others [42–46].

This work builds on this latter approach to capitalize on the ranking property of this spectral technique. We exploit the feedback qualities captured by node centrality measures such as the PageRank [47] algorithm and Bonacich's eigenvector centrality [48]. We exploit the notion in which nodes are central either by having many connections, or by being well connected to other central nodes to unveil new central places that by simple measures of population density would not be considered as such.

**METHODS**

**Mobility Data Aggregation**

Mobility data for a set of eleven urban regions in Chile were compiled from the Origin-Destination surveys (OD hereafter) conducted by the Department of Transportation and Planning (SECTRA in Spanish) at the Chilean Ministry of Transportation and Telecommunications[49] between the years 2010-2014. Analyzed regions are: Arica, Antofagasta, Iquique-Alto Hospicio, Copiapó, La Serena-Coquimbo, Valparaíso-Viña del Mar,

Santiago, Temuco, Valdivia and Osorno, covering most regional capitals and metropolitan areas in Chile.

We follow OD zones defined by SECTRA as basic geographic units within each OD area for the analysis. Each OD survey reports the number of trips between OD zones in a typical labor day for every polled individual within urban areas. An expansion factor, calculated by SECTRA from population projections, is used to estimate the total number of trips and population size within a particular OD zone (Table 1).

*Table 1. Summary statistics of analyzed OD surveys. Total population and number of trips are estimated using the expansion factor computed by SECTRA. The actual number of surveyed subjects and trips are shown in parenthesis.*

| OD SURVEY (ID) | POPULATION | TRIPS | YEAR |
|---|---|---|---|
| ARICA (A) | 193,073 (6,189) | 568,053 (18,417) | 2010 |
| IQUIQUE - ALTO HOSPICIO (B) | 267,887 (9,014) | 653,181 (21,248) | 2010 |
| ANTOFAGASTA (C) | 329,294 (9,505) | 831,484 (23,789) | 2010 |
| COPIAPÓ (D) | 145,683 (6,197) | 417,876 (17,247) | 2010 |
| LA SERENA - COQUIMBO (E) | 366,463 (10,687) | 928,209 (27,157) | 2010 |
| VALPARAÍSO - VIÑA DEL MAR (F) | 964,565 (27,504) | 2,295,100 (52,726) | 2014 |
| GRAN SANTIAGO (G) | 6,651,735 (60,054) | 18,461,134 (78,820) | 2012 |
| TEMUCO (H) | 311,873 (10,073) | 1,008,087 (26,668) | 2013 |
| VALDIVIA (I) | 161,245 (6,931) | 561,830 (18,646) | 2013 |
| OSORNO (J) | 138,967 (6,647) | 468,652 (16,804) | 2013 |

## Spectral Analysis of the Human Mobility Networks

Undirected networks are built for each OD survey. In these networks, nodes represent OD zones connected by an edge if, at least, one trip is recorded between nodes *i* and *j*. We weigh edges by the total number of trips $T_{ij}$ between zones after multiplying by the corresponding expansion factor. We include self-loops (i.e. trips within zones).

Each network, B, is represented by $B_{ij} = T_{ij} - P_{ij}$, an adaptation of the *modularity matrix* proposed by Newman[41,50], where $P_{ij}$ is the null model that represents the expected number of edges between two given nodes in a random network with the same degree distribution of the empirical network. In this particular null model [41,51], $P_{ij}$ is defined as $P_{ij}=k_i*k_j/2m$, where $k_i$ is the degree of node *i*. For the mobility networks analyzed here, the degree $k_i$ represents the sum of both incoming and outgoing trips at a given OD zone, equivalent to the generalization of the null model for weighted networks [51].

## Place Ranking and Scaling Laws

We employ the leading eigenvalue ($\lambda$) of **B** and its corresponding eigenvector, $x$, as a measure of place influence for each OD zone, equivalent to the measure of node importance known as eigenvector centrality [49] $\psi_i = \|\lambda x_i\| = \|\sum_j B_{ij} x_j\|$, where $x_j$ represents the centrality of node *j*. As the *modularity matrix* **B** can contain negative values, we consider the absolute value of this centrality index. Likewise, we present the eigenvector centrality measure in its non-normalized form. Since $\psi_i$ represents the convergence of $B^k$ to the ratio of its components, it is an approximate measure of the number of trips captured by each OD zone, either directly or indirectly. The expansion factor for each trip, allows the comparison of such centrality measure across the full set of OD surveys from which we construct a ranking of OD zones at the national level, according to their relevance in the underlying mobility network.

The centrality $\psi_i$ also captures a feedback quality in the network, where OD zones are central not only by generating or receiving many trips, but also by being well connected to other central OD zones. We posit that this measure of centrality allows the identification of places such as, for example, low-density suburbs, bedroom communities, or "leap-frog" developments (e.g. urban development detached from the main urban core) that by measures of physical concentration (i.e. population or trip density) would typically fail to be considered as central.

The regions covered by each OD survey are then classified in two clusters, comprised by OD zones, dissecting what we nominate urban areas (U) from the rural hinterland (R). Additionally, we evaluate the scaling relationship between the number of trips, $T_i$, and population, $P_i$, based on their eigenvector centrality $\psi_i$. We follow the procedure outlined in [35,52], and simulate a one-dimensional parameter space representing thresholds $\psi^n$ from the log-normal distribution of centrality scores $\psi_i$, derived for the full set of OD zones.

We divide each survey in two clusters containing the OD zones with centrality values above and below each simulated threshold $\psi^n$. These divisions represent several possible pairs of U and R areas where the scaling relationship between the total population size (P) and total number of trips (T) was evaluated. Note that outbound trips of the surveyed areas in the original dataset were not aggregated down to the OD zone level and thus were not accounted in this analysis, as well as external OD zones where no inhabitants were surveyed. We then fit a power law using the linearization: $log_{10}(T) = log_{10}(T_0) + \beta\ log_{10}(R)$. At each centrality threshold $\psi^n$, we derive two distinct scaling regimes for U and R clusters, across the full set of OD surveys.

**RESULTS**

Scaling between *T* and *P* is sublinear ($\beta$ = 0.93) prior to clustering (Figure 1). Figure 2 shows that, after clustering, two distinct scaling regimes seem to hold throughout most urban delineations. We disregard the first and last thresholds below $10^2$ and above $10^3$ as their high variability is most likely associated to the small sample sizes of small R clusters in the lower bound and almost negligible population in the upper bound of $\psi^n$. Two distinct scaling regimes may be identified within these boundaries. U clusters (i.e. with large centrality OD zones) show a consistently higher scaling exponent compared to R clusters based on their non-overlapping 95% confidence intervals.

*Figure 1. Regression of Total Population against Total Number of Trips, estimated from Origin-Destination (OD) surveys of ten cities in Chile. Letters correspond to each urban area analyzed (see Figure 3). Regression model: log10(T)=log10(0.83) + 0.95 log10(P), R2=0.98, C.I. [0.83,1.04].*

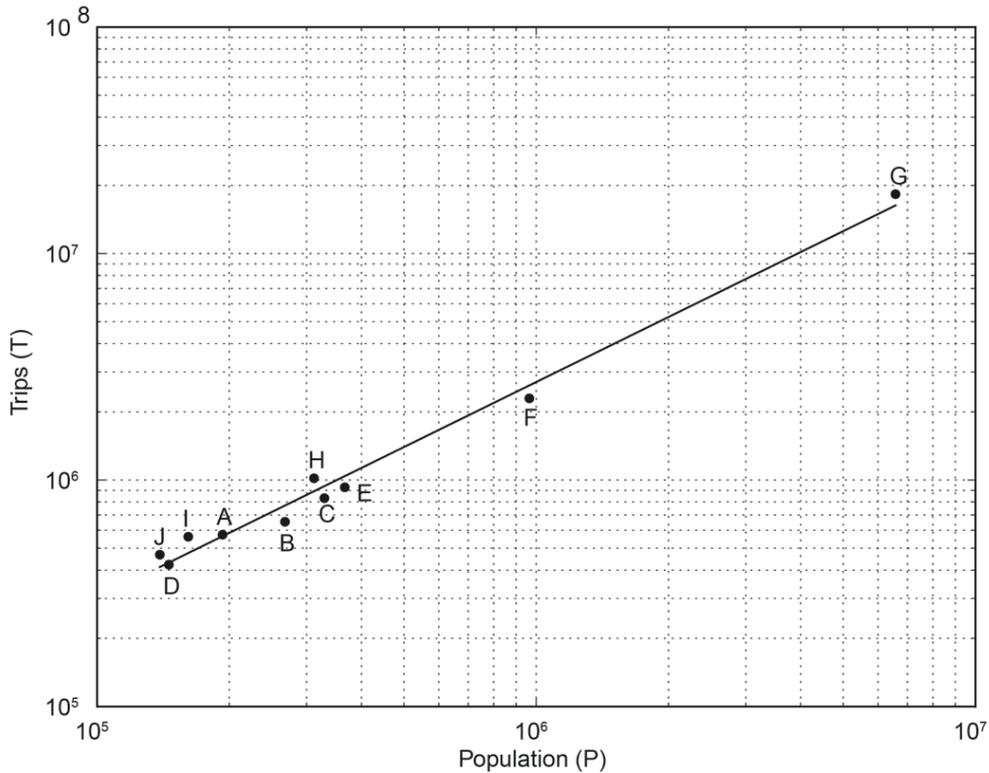

From Figure 2, it is possible to identify two qualitative thresholds of particular interest (see also Tables 2 and 3 and Figure 2). The first at $\psi^a \approx 10^{2.14}$, represents 138 trips over the expected null model. At this first threshold, the OD zones identified as rural exhibit a sub-linear scaling regime ($\beta$ = 0.70 ± 0.16, Figure 4), while urban clusters show a close-to-linear scaling regime ($\beta$ = 0.95 ± 0.14, Figure 4). Note also that population estimates within these geographic boundaries are fairly close to official estimates (Table 2) according to city definitions used by the Chilean National Statistics Agency (INE) and the Chilean Housing Ministry (MINVU)[54].

*Figure 2. Sensitivity analysis of scaling exponent for different delineations of urban (U) and rural (R) areas. At each centrality threshold ψ n, OD surveys are divided into U areas with high centrality values in blue, and the remaining R clusters in green. Scaling exponents β, are plotted with their 95% confidence intervals. The horizontal dotted line represents the baseline situation (i.e. Figure 1), while the dotted vertical lines (ψa and ψb) represent the thresholds of interest.*

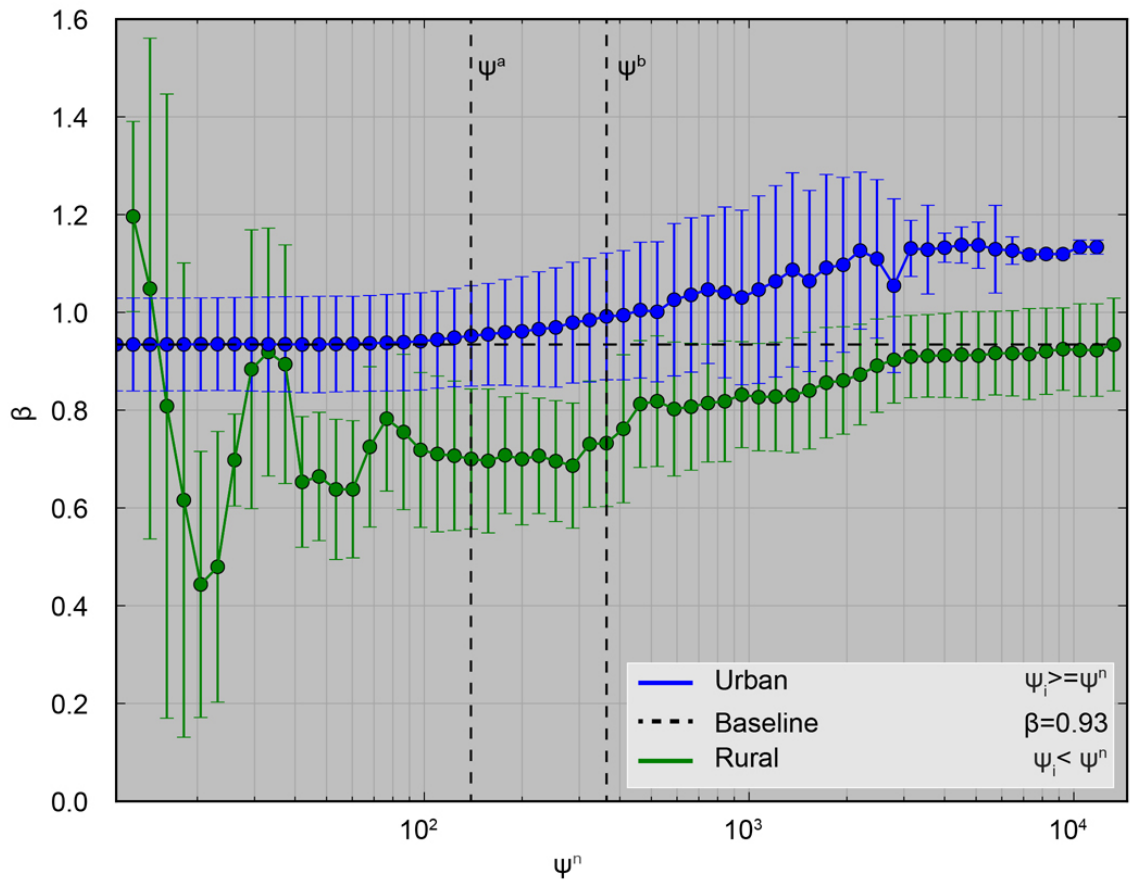

The second threshold arises at $\psi^b \approx 10^{2.56}$, however, it seems far too stringent to be employed to define city boundaries as it would lead to unrealistic representations of current urban population compared to their rural counterpart (Tables 2 and 3). For instance, choosing $\psi^b$ as the urban-rural threshold would imply having a ~50% (i.e. ~$3 \times 10^6$ inhabitants) of rural population in the greater Santiago area, which clearly defies the urban definition in a city of ~$5.5 \times 10^6$ people (see Table 2). Another salient feature emerging from this analysis, is that scaling exponents for U areas are surprisingly similar to the general scaling of trips across OD areas shown in Figure 1 (see Table 3), while U areas remain distinctly sub-linear.

Table 2: Population size of defined urban and rural aggregates as estimated from the eigenvector centrality thresholds for each OD survey. Last column includes the population estimates for urban areas from the Chilean Housing and Urbanism Ministry, based on 2002 Census Data.

| | Eigenvector centrality thresholds | | | | Population |
|---|---|---|---|---|---|
| | $\psi^a$ | | $\psi^b$ | | |
| OD SURVEY (ID) | Rural | Urban | Rural | Urban | |
| ARICA (A) | 1,374 | 147,583 | 4,413 | 144,544 | 175,441 |
| IQUIQUE-ALTO HOSPICIO (B) | 820 | 248,652 | 2,580 | 246,892 | 214,586 |
| ANTOFAGASTA (C) | 16,769 | 312,564 | 81,921 | 247,412 | 285,255 |
| COPIAPÓ (D) | 8,149 | 136,772 | 37,169 | 107,752 | 134,561 |
| LA SERENA-COQUIMBO (E) | 10,668 | 336,426 | 35,886 | 311,208 | 296,253 |
| VALPARAÍSO-VIÑA DEL MAR (F) | 81,768 | 879,673 | 217,235 | 744,206 | 824,006 |
| GRAN SANTIAGO (G) | 849,451 | 5,724,438 | 3,120,864 | 3,453,025 | 5,631,839 |
| TEMUCO (H) | 17,043 | 279,549 | 103,269 | 193,323 | 268,437 |
| VALDIVIA (I) | 11,978 | 149,043 | 36,483 | 124,538 | 127,750 |
| OSORNO (J) | 3,757 | 116,851 | 7,654 | 112,954 | 132,245 |
| TOTAL | 1,001,777 | 8,331,551 | 3,647,474 | 5,658,854 | 8,090,373 |

Table 3: OLS fit summary among cluster realizations for cities. Urban and rural areas were defined based upon thresholds (see Figure 2)

| | Eigenvector centrality thresholds | | | |
|---|---|---|---|---|
| | $\psi^a$ | | $\psi^b$ | |
| | Rural | Urban | Rural | Urban |
| Slope ($\beta$) | 0.70 | 0.95 | 0.73 | 0.99 |
| CI$_\beta$ | (0.54, 0.86) | (0.84, 1.06) | (0.58, 0.87) | (0.85, 1.14) |
| Intercept | 2.11 | 0.75 | 2.06 | 0.57 |
| Adj. $R^2$ | 0.96 | 0.99 | 0.98 | 0.99 |

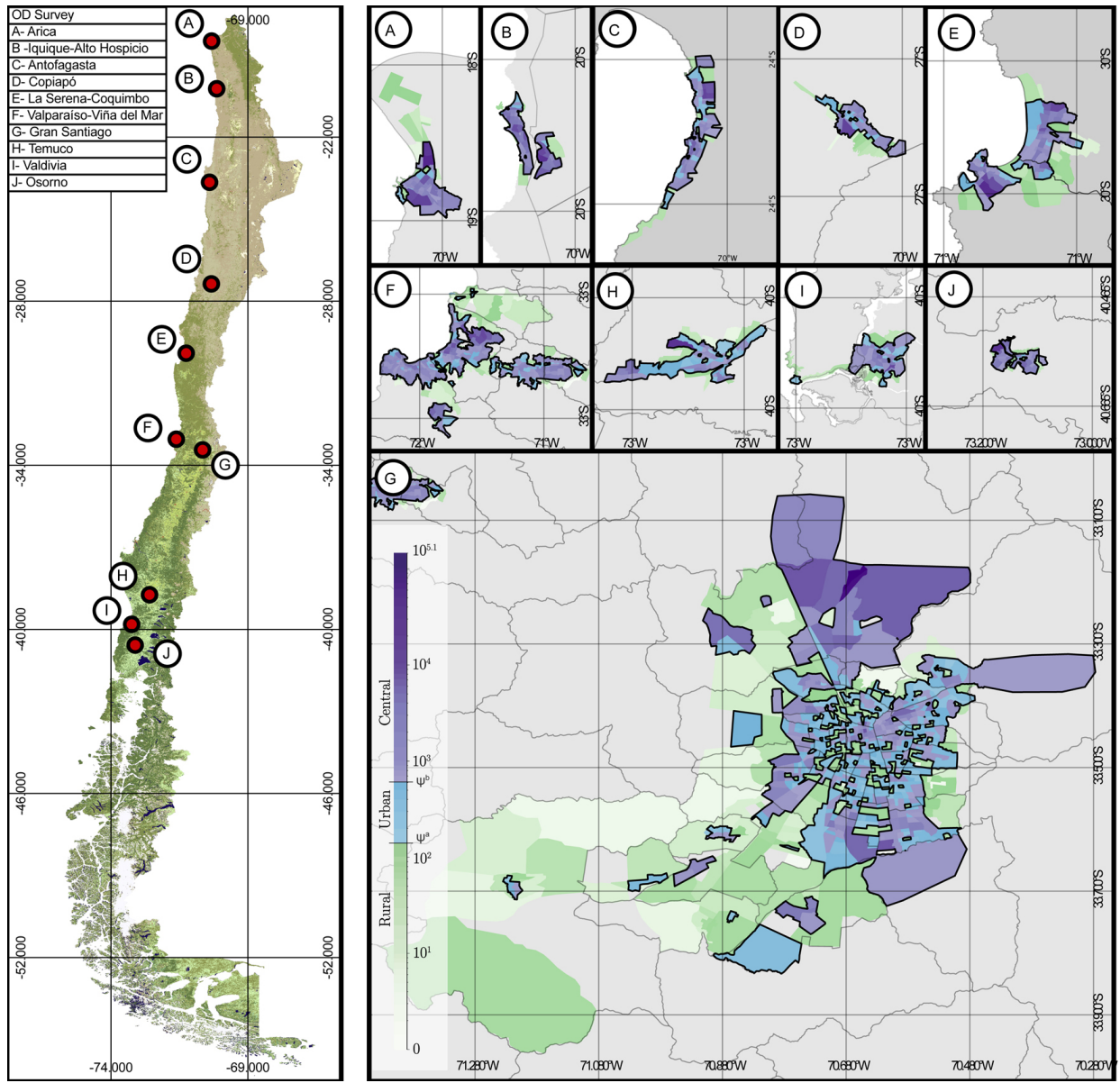

Figure 3. OD survey regions and OD zones by threshold values. Shades of greens indicate OD zones branded as "rural" areas, while shades of blue and violet indicate OD zones belonging to "urban" and "central" areas respectively.

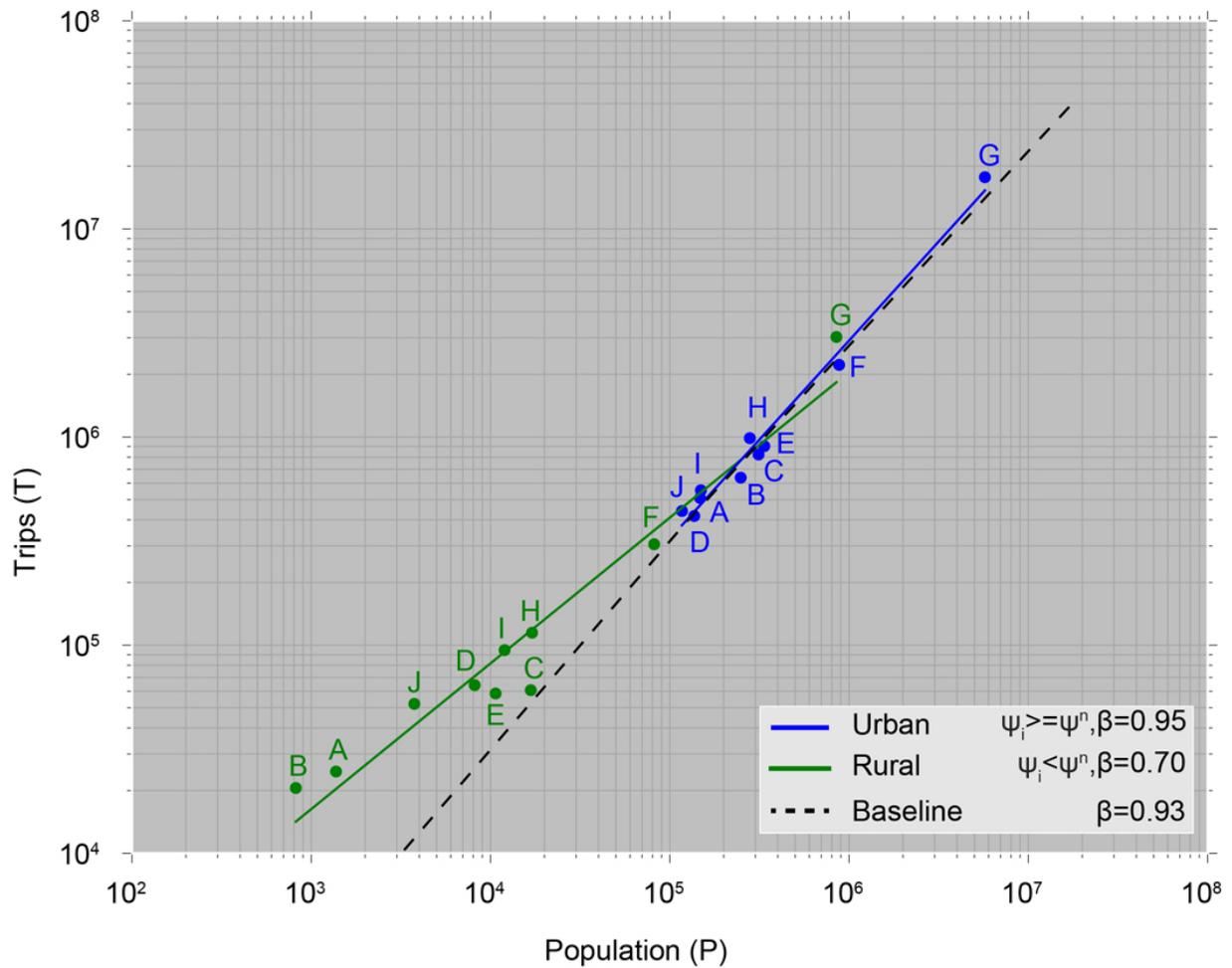

*Figure 4. Total population against total number of trips across OD surveys, for urban (blue) and rural (green) areas as defined by centrality threshold ψ a=102.14.*

**DISCUSSION**

Based on the premise that urban scaling exponents may be an expression of how underlying infrastructure shape human interactions, we here contribute towards setting an explicit link between urban boundaries and urban scaling laws. We model cities by ranking places as a function of their centrality, which may also be understood as their influence in the underlying mobility network. Such approach has the advantage to admit the evaluation of changes in the scaling exponent at different urban extents given by our delineation process, thereby expanding on the theoretical frameworks contributed by Bettencourt et al. [20], Arcaute et al.

[35], Ratti et al. [42], Cottineau et al. [8],Martinez [56], Barthelemy [18]., Hanley et al. [52] and others in the emerging domain of the *New Science of Cities* [19].

For the mobility dataset analyzed here, we hypothesize that, if divided into proper urban and rural areas, regions should show consistent differences in their scaling regimes of urban outputs. The dissection of regions, in what we have termed as urban and rural areas, shows that coarse and nonfunctional definitions of cities could lead to the grouping of places with noticeably distinct social and economic dynamics. In fact, if urban boundaries are defined using the first threshold defined here, we may define urban areas when OD zones exceed the 138 trips threshold (i.e. $\psi^a \approx 10^{2.14}$) which coincides with the extent of impervious surface, as derived in recently developed land-cover classification by Zhao et al. [56] (Figure S.1 in Supplementary Materials).

We expect that these results will contribute to stress the relevance of properly defining cities from a functional standpoint, which seems to constitute a crucial piece in a theory linking micro dynamics of social interaction with emerging properties of human aggregates [25,35]. The analysis presented here sheds additional light on the inconvenience of relying on simple city definitions solely based on administrative divisions or spatial concentration criteria, as these tend to delineate coarse regions where urban places and their rural hinterland could be lumped together.

While our evaluation of scaling relationships suggests that, for the case of human mobility, urban and rural places cannot be safely assumed to be of equal quality [57], we nevertheless believe that this particular analysis calls for a trial of the method on more comprehensive mobility datasets. The dataset analyzed here is limited in its coverage, lacking data about smaller cities, which would provide valuable information on the boundaries on the tail of the distribution [54].

The study of trip length scaling has been an area of active research in the past decade [34,59–60], though to our knowledge no such analysis has been undertaken across urban systems. Naturally, this kind of analysis requires non-trivial definitions, such as how to define a proper trip, and the convenience of abstracting trips from their length. However, we posit that using scaling as a diagnostic will contribute to a better understanding of urban dynamics. Hence, resolving how scaling of urban trips in isolation deserves further study, as the question of whether larger cities produce more trips per-capita is indeed relevant for understanding congestion and planning for ever growing cities. Recent attempts to provide first principle explanations to urban scaling show urban scaling as emerging from cost-effective mobility at lower levels that sometimes considers the cost of transportation given the underlying network of connected infrastructure [18,54,55]. Beyond ongoing debate regarding the correct approach to evaluate scaling [25,58], we consider that it would be of interest to clarify if economies of scale emerge in larger cities due to increased mobility (which would imply superlinear scaling of trips), or by the possibility of reaching more extended areas (consistent with near-to-linear scaling of trips). So far, our results contribute to explicitly show that rural places and their inhabitants are relatively disconnected from other places when compared to urban areas, in addition to provide a methodology to geographically delineate where this happens.

Our functional definition of urban boundaries could aid in bridging the problem of city definition with theoretical descriptions of the urban environment akin to the notion of cities as social reactors [63]. This could not only be of potential use for researchers, but also for planners and policy makers, as this methodological proposition delineates urban areas from transportation surveys. The ranking of places in accordance to the spectral decomposition of their underlying mobility network still leaves ample room to further explore the clustering properties of these techniques, which have been explored in several applications of geographic segmentation, such as the redefinition of administrative boundaries from phone calls [42–45], the geographic clustering of cities from credit card transactions [46], detection of land uses from mobile phone records[64], and the detection of urban mega-regions from mobility surveys [8,35,65,66].

**Acknowledgments**

Funding for this research was provided by FONDEF-CONICYT grant # ID15I10313 and FONDECYT-CONICYT grant #1161280

# Supplementary Materials

## Qualitative assessment of threshold delineation using satellite land cover data.

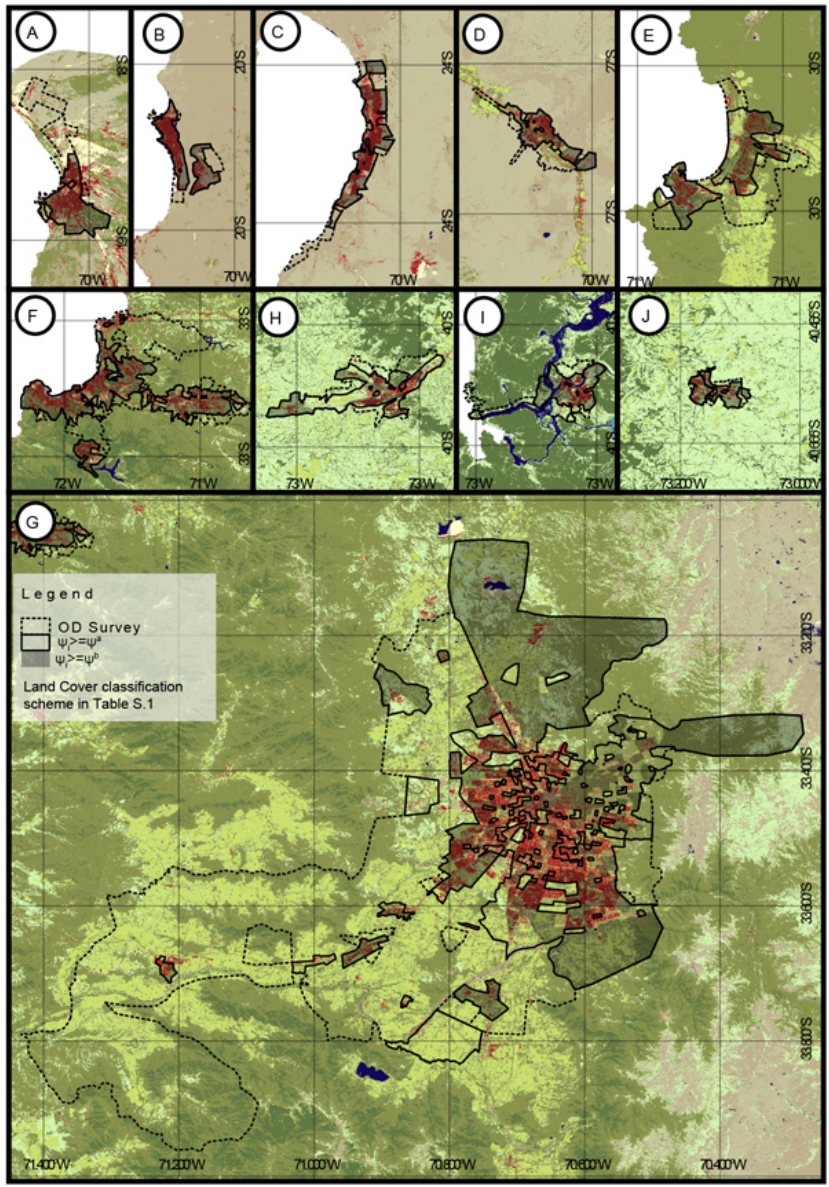

*Figure S.1. Comparison of city delineations as defined by thresholds of interest $\psi^a$ and $\psi^b$ with Land Cover raster dataset derived from remote sensing. In dotted lines are shown the original extents of the areas covered by the OD Surveys (excluding the areas branded as "External", which were not taken into account in the analysis, see Main Text). Color codes for the 30 m x 30 m pixels represented in the Land Cover are included in Table S.2.*

*Table S.1: Classification and Color Scheme for the Land Cover dataset derived from remote sensing. Original codes are in parenthesis [1]. The last column shows color codes for the 30 m x 30 m pixels displayed in Figure S.3.*

| Level 1 | Level 2 | Level 3 | |
|---|---|---|---|
| **Clouds 120 (1200)** | | | |
| **Ice and Snow 200 (1000)** | 210 Snow (1010) | | |
| | Ice 220 (1020) | | |
| **Water 300 (600)** | Lakes 310 (610) | | |
| | Reservoirs 320 (620) | | |
| | Rivers 330 (630) | | |
| | Ocean 340 (640) | | |
| **Wetland 400 (500)** | Marshes 410 (510) | | |
| | Swamps 420 (520) | | |
| | Others 430 (530) | | |
| **Grasslands 500 (300)** | Meadow 510 (310) | Seasonal 510 (310) | |
| | | Evergreen 520 (320) | |
| | Others 520 (320) | | |
| | Arid Grasslands 530 (330) | | |
| **Forests 600 (200)** | Native Broadleaf 610 (210) | Primary 611 (211) | |
| | | Renewable 612 (212) | |
| | Native Coniferous 620 (220) | Primary 621 (221) | |
| | | Renewable 622 (222) | |
| | Mixed 630 (230) | Primary 631 (231) | |
| | | Renewable 632 (232) | |
| | Broadleaf Plantations 640 (240) | Mature 641 (241) | |
| | | Harvest 642 (242) | |
| | Coniferous Plantations 650 (250) | Mature 651 (251) | |
| | | Harvest 652 (252) | |
| **Crops 700 (100)** | Rice 710 (110) | | |
| | Greenhouses 720 (120) | | |
| | Others 730 (130) | | |
| | Orchards 740 (140) | | |
| | Fallow 750 (150) | | |

| Shrubs 800 (400) | Shrubs 810 (410) | |
| --- | --- | --- |
| | Tree-like shrubs 820 (420) | |
| | Succulents 830 (430) | |
| | Shrub Plantations 840 (440) | |
| | Other Arid Shrubs 850 (450) | |
| **Barren 900 (900)** | Salt Flats 910 (910) | |
| | Sands 920 (920) | |
| | Rocks 930 (930) | Rocks 931 (931) |
| | | Gravel 932 (932) |
| **Impervious Surface 1000 (800)** | | |

## Sources